\documentclass{blois}

\def\Journal#1#2#3#4{{#1} {\bf #2}, #3 (#4)}

\def\be{\begin{equation}}
\def\ee{\end{equation}}
\def\bea{\begin{eqnarray}}
\def\eea{\end{eqnarray}}

\newcommand{\Photo}{\includegraphics[height=35mm]{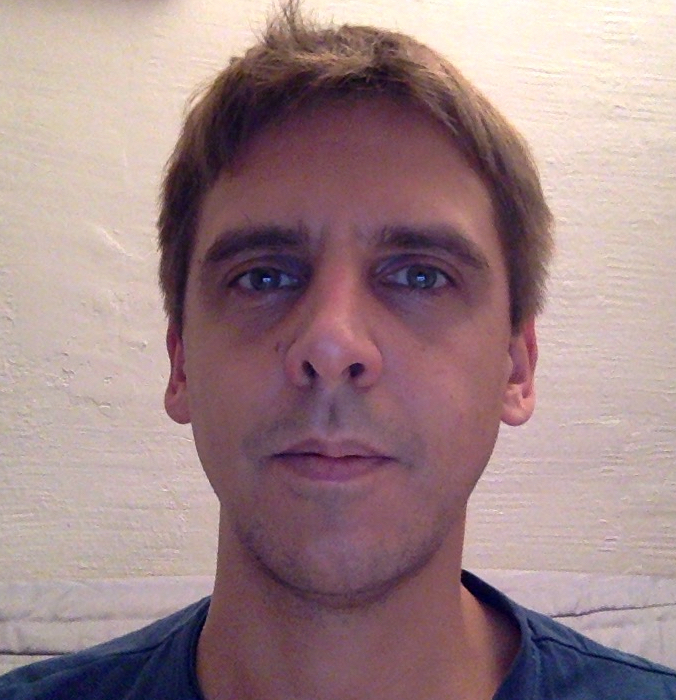}}

\begin{document}
\vspace*{4cm}
\title{NEWS FROM THE VERY-HIGH-ENERGY SKY SEEN WITH H.E.S.S.}

\author{ A. ZECH on behalf of the H.E.S.S. COLLABORATION }

\address{LUTH, Observatoire de Paris, CNRS, PSL, Universit\'e de Paris\\
 5 pl. Jules Janssen, 92195 Meudon, France \vspace{0.3cm}   }

\maketitle\abstracts{
The H.E.S.S. experiment, the largest Cherenkov telescope array 
to date, has been observing the sky at TeV energies for the past 
16 years. Its location in the Southern hemisphere provides H.E.S.S. 
with equally good access to Galactic and extra-galactic sources. 
The focus of observations is now gradually shifting from discoveries 
of new TeV emitters to studies of transients and the detailed exploration 
of known sources. A few recent highlights from these observations are 
presented here, with a focus on extragalactic observations. }

\section{The H.E.S.S. array in 2019}
The array of the High Energy Stereoscopic System, located in Namibia in the Khomas Highland, consists
of five Cherenkov telescopes of two different types. Four 12\,m telescopes of Davies-Cotton design with a field
of view of 5$^{\circ}$ have been fully operational since 2004. In 2017, their sensitivity has been improved
thanks to an upgrade of their cameras --- equipped with photomultiplier tubes --- with the "NectarCAM" electronics developed 
for the Cherenkov Telescope Array (CTA).
A fifth, parabolic telescope with a diameter of 28\,m ("CT5") and a field of view of 3.5$^{\circ}$, was added to the array in 
2012. Observations in stereoscopic mode, i.e.\ combining at least two out of the five telescopes, have an energy threshold 
of about 80\,GeV, while monoscopic observations with the CT5 telescope alone achieve a very low threshold of about 30\,GeV, 
at the expense of a lower overall sensitivity. 

While the very-high-energy (VHE) community is preparing the installation of CTA, H.E.S.S. is at present the only experiment 
combining different types of Cherenkov telescopes in a single array. It is also the only array in the Southern hemisphere, which 
provides it with a privileged view of the Galactic Plane and Center and of certain particularly interesting extragalactic sources, 
like the most nearby radio galaxy Centaurus\,A.

\section{News from the Galactic sky}

\subsection{The H.E.S.S. Galactic Plane Survey.}
One of the most visible achievements of the "phase-I" observations with H.E.S.S. (including only the four 12\,m telescopes), 
published in 2018, is the 2700\,hr long survey of the Galactic Plane (HGPS), covering about 180 degrees in longitude and $\pm$3
degrees in latitude above and below the plane, with an angular resolution of 0.08$^{\circ}$, resulting in a sensitivity better 
than 2\,\% of the flux from the Crab nebula~\cite{hgps}. The catalog includes also sources from additional Galactic regions, e.g.\
around the Crab nebula or the supernova remnant SN\,1006.
This survey has led to the discovery of a total number of 78 sources. Among the firmly identified sources are 12 pulsar 
wind nebulae (PWN), 8 supernova remnants (SNR), 8 composite PWN/SNR sources, and 3 $\gamma$-ray binaires.

The number of the latter has increased in the meantime, thanks to more recent observations including also CT5, which led to the discovery of VHE 
emission from Eta Carinae~\cite{eta} and LMC\,P3~\cite{lmcp3}. 
In addition, the lowered energy threshold has permitted the detection of pulsed emission from the Vela pulsar~\cite{vela} and from PSR\,B1706-44~\cite{psr}.
Eleven sources in the catalog are not clearly associated with counterparts at other wavelengths and 36 
sources are not firmly identified, i.e.\ often they can be ascribed to more than one counterpart. This last category includes most likely a large number of PWN.

\subsection{The extreme accelerator HESS\,J1826-130}
One example of a very interesting source in the Galactic Plane that was initially discovered during the survey, but required twice as many hours
of observations for a full spectral and morphological characterisation, is the extended source HESS\,J1826-130. The analysis of this still unidentified object 
was complicated by the unusual hardness of its spectrum and by contamination from a bright nearby PWN (HESS\,J1826-137). A preliminary spectral analysis yields a power-law fit of $dN/dE \propto E^{-1.78}$ and a cutoff at about 15\,TeV~\cite{1826}. This makes it one of the hardest spectra ever observed with H.E.S.S. Two scenarios have been proposed to explain the nature of this source. Its emission might be due to Inverse Compton radiation from electrons and positrons 
from the Eel nebula (a PWN consistent with the location of the H.E.S.S. source) interacting with ambient photons. A more intriguing scenario ascribes the emission to pion production of highly energetic cosmic rays from the progenitor SNR of the nearby bright PWN on molecular clouds 
in the field of view of HESS\,J1826-130. This would require protons with an energy of a few 100\,TeV, making this source a very promising PeVatron candidate.

\section{News from the extragalactic sky}

\subsection{The H.E.S.S.-I legacy extragalactic survey}
A considerable amount of observation time is also dedicated to observations of extragalactic sources. About 2600\,hr of "phase-I" observations up to 2013, corresponding 
to a coverage of more than 6\% of the extragalactic sky, have now been re-analysed with the most recent data reduction methods~\cite{egal}.
Twenty-four sources were re-detected, six out of which have proven to be significantly variable, following a blind search for transient or variable sources. 
The HESS Phase-I extragalactic source population is largely dominated by active galactic nuclei (AGN) of the blazar type : 
one flat-spectrum radio quasar (FSRQ), one low frequency peaked BL Lac objects (LBL), 15 high frequency peaked
BL Lac objects (HBL) and three ultra-high-frequency peaked BL Lacs (UHBL or EHBL). Three radio galaxies and one starburst galaxy, NGC\,253, complete the
count. With more recent observations including CT5, two further FSRQs, PKS\,0736+017 and 3C\,279, and an additional LBL, PKS\,1749+096, have been detected. 
Target of Opportunity (ToO) observations, described below, have now increased the count of variable sources to ten.

The combined extragalactic field of view covered by the H.E.S.S.-I data set contains 184 objects from the {\it Fermi} 3FGL point-source catalog~\cite{3fgl}. 
Upper flux limits in the VHE range can be derived for these objects from the H.E.S.S. data in case of non-detection. The extrapolated 
{\it Fermi}-LAT spectrum, after absorption with a model of the extragalactic background light (EBL), can be compared to the H.E.S.S. spectrum or upper flux limits. 
For eleven 3FGL objects, such comparisons resulted in H.E.S.S. upper limits below the predicted flux from extrapolations, providing proof of intrinsic spectral turnovers in these sources. The extragalactic source catalog, together with sky maps, will be released in the near future.

\subsection{News from Centaurus\,A}
One of the most prominent extragalactic sources on the Southern sky is the most nearby radio galaxy Centaurus\,A (Cen\,A), at a distance of
3.8\,Mpc, with its large radio lobes covering about 8$^{\circ}$ on the sky. Its core region was first detected with H.E.S.S. in 2009,
thanks to a cumulation of more than 120\,hr of observations. A recent joint analysis of 213\,hr of H.E.S.S. data and a {\it Fermi}-LAT data 
set spanning 8 years of observations confirmed the existence of a spectral hardening around a few GeV~\cite{cena}. 
Following the general paradigm that FR-I type radio galaxies, like Cen\,A, are the parent population of BL Lac type blazars, one would
naively expect a similar high-energy emission for such sources, without any spectral features at the highest energies.
The observed spectral upturn indicates that the emission cannot be simply ascribed to the usual one-zone synchrotron self-Compton scenario 
that is very successfully applied to many BL Lacs. A more complicated scenario, either requiring several populations of 
emitting particles or several emission regions, needs to be explored. 

Another very recent result confirms now the insufficiency of a scenario where all the high-energy emission would come from a single
compact region. H.E.S.S. has detected extended VHE emission from Cen\,A that is aligned
with the kpc-scale jets of the source. The emission above a few hundred GeV is consistent with an extension exceeding 1.8\,kpc. This is 
the first time that extended VHE emission has been detected from an extragalactic source. One explanation,
consistent with an earlier prediction~\cite{cenasim}, supposes that Inverse Compton up-scattering of mostly starlight by highly relativistic 
electrons along the jets is at the origin of the observed emission. This would of course require an efficient mechanism to compensate for cooling 
processes by re-accelerating those particles over large distances to very high energies. More sophisticated models, combining the macrophysics 
of the relativistic jets with the microphysics of particle acceleration and emission, will be needed to fully understand the available data. Future 
observations with CTA for this source have already been foreseen as part of its Key Science Projects~\cite{ctaksp}.

\subsection{News from transient sources}
In addition to deep observations of nearby steady sources, follow-up observations of multi-wavelength (MWL) and multi-messenger alerts have
taken an increasingly important place in the H.E.S.S. schedule. Well sampled information on the variability of Galactic and extragalactic sources adds 
an additional observable that is crucial to understanding the underlying mechanisms in VHE emitting objects. 
%Galactic ToO observations have recently led to the measurement of upper flux limits on 10 core-collapse supernovae, from which upper limits were derived on the mass loss rates of their progenitor stars~\cite{sn}.
Extragalactic targets are mostly AGN flares and gamma-ray bursts (GRBs) and, to a lesser extent, fast radio bursts. The low energy threshold achievable with the CT5 telescope is playing a major role in such observations, since flaring sources are often relatively distant and strongly absorbed by the EBL at TeV energies. 

In the first half of this year, 10 different AGN were observed following MWL alerts. The most spectacular recent flares where detected from the FSRQ 3C\,279. Data from a MWL follow-up of a bright flare in June 2015 permitted to strongly constrain the location of the emission zone to beyond the broad line region. Intriguingly, neither a standard leptonic, nor a lepto-hadronic scenario based on a single emission region are able to satisfactorily reproduce the full flare data set~\cite{3c279}. Recent data from two flares detected in 2018 are under analysis.

The regular follow-up of GCN alerts to GRBs has finally resulted in its first discovery of such sources for H.E.S.S. Surprisingly, VHE emission above 100\,GeV from the long GRB\,180720B was still detected roughly 10 hours after the outburst. By the time of writing, a second, very nearby GRB (190829A) has been detected~\cite{grb19}, improving further the very promising outlook for future detections with CTA. 

The H.E.S.S. array was the first one to follow up on the IceCube alert of the high-energy neutrino that has been 
associated with the blazar TXS\,0506+056. Unfortunately, 1.3 hours of observations starting about 4\,hr after the alert did not reveal a
significant signal. Later detections with the MAGIC and VERITAS telescopes seem to indicate a delay of the VHE emission with 
respect to the neutrino event~\cite{txs}. Attempts at modelling simultaneously the MWL and neutrino emission from this source have proven difficult~\cite{keivani2018,cerruti2019,reimer2019,gao2019}. 

H.E.S.S. ToO observations also covered several gravitational wave alerts from the Advanced LIGO/VIRGO 01 run and 
provided an upper limit on the VHE flux from the neutron star merger event GW\,170817~\cite{GW}. Follow-up observations of the 
ongoing 03 run are under way. 

\section {The future of the H.E.S.S. array}

The H.E.S.S. experiment has been prolonged for at least three years, until the end of 2022. 
During the remaining years, the focus will be on the scientific exploitation on the one hand and
on prototyping for CTA on the other. ToO observations and multi-wavelength campaigns will continue 
to play a major role. An upgrade of the CT5 camera electronics with a CTA type "FlashCAM" will be carried 
out before the end of this year, providing improved performance and an opportunity to test 
an important component of the future observatory in the field.

A first limited set of raw data and instrument response functions for four Galactic and extragalactic sources
has been made publicly available for the first time~\cite{release}. This allows
the wider community to get accustomed to the future CTA data format and to test the software and analysis methods
that are currently being developed. Further releases are envisaged.

\section*{Acknowledgments}

\footnotesize{
\url{https://www.mpi-hd.mpg.de/hfm/HESS/pages/publications/auxiliary/HESS-Acknowledgements-2019.html}
}

\normalsize

\section*{References}

\small{

}
 
\end{document}